\documentclass[sigconf]{acmart} 

\acmConference[ABC]{ABC}{2023}{USA}

\usepackage[ruled, lined, linesnumbered, commentsnumbered, longend]{algorithm2e}

\begin{document}

{\renewcommand{\baselinestretch}{1}}
\title{ZeRO++: Extremely Efficient Collective Communication for Giant Model Training}

\author{Guanhua Wang*, Heyang Qin*, Sam Ade Jacobs, Connor Holmes, Samyam Rajbhandari}
\author{Olatunji Ruwase, Feng Yan$^1$, Lei Yang$^2$, Yuxiong He}
\affiliation{
  \institution{Microsoft}
  \country{}}


\affiliation{ \textit{\{ guanhuawang, heyangqin, samjacobs, connorholmes, samyamr, olruwase, yuxhe \} @microsoft.com}
\country{}}

\begin{abstract}
 Zero Redundancy Optimizer (ZeRO) has been used to train a wide range of large language models on massive GPUs clusters due to its ease of use, efficiency, and good scalability. However, when training on low-bandwidth clusters, or at scale which forces batch size per GPU to be small, ZeRO's effective throughput is limited because of high communication volume from gathering weights in forward pass, backward pass, and averaging gradients. This paper introduces three communication volume reduction techniques, which we collectively refer to as ZeRO++, targeting each of the communication collectives in ZeRO. First is block-quantization based all-gather. Second is data remapping that trades-off communication for more memory. Third is a novel all-to-all based quantized gradient averaging paradigm as replacement of  reduce-scatter collective, which preserves accuracy despite communicating low precision data. Collectively, ZeRO++ reduces communication volume of ZeRO by 4x, enabling up to 2.16x better throughput at 384 GPU scale.
\end{abstract}



\maketitle


\section{Extended Introduction}
\label{sec:intro}

Deep learning (DL) models have been applied successfully in many different domains such as image/video analysis, natural language processing, speech recognition, etc. Over years, the quality, functionality, and coverage of these models have continued to improve. Model size has been a key factor in this improvement. There is a strong correlation of model size with accuracy and improved functionality, and as result, the model size has grown dramatically in recent years. For example, parameter size grows from 100 million to 500+ billion from  BERT \cite{devlin2018bert} to Megatron-Turing NLG \cite{smith2022using}. 

With the increase in model size, the memory and compute requirements for training have increased significantly beyond the capability of a single accelerator (e.g., a GPU). Training massive models requires efficiently using aggregated computing power and memory across hundreds or even thousands of GPU devices. There are two popular approaches to this, namely 3D parallelism \cite{megatron-2021,deepspeed3dblog} and Zero Redundancy Optimizer (ZeRO) \cite{rajbhandari2020zero}. 

3D parallelism combines data parallelism \cite{ben2019demystifying,dean2012large},  pipeline parallelism \cite{narayanan2021memory,harlap2018pipedream,huang2019gpipe} and tensor parallelism \cite{shoeybi2019megatron} to distribute model training workloads across hundreds of GPUs. This approach can achieve excellent per-GPU computing and memory efficiency. However, a major downside here is the system and user complexity. It puts the burden of refactoring the single GPU code to work for 3D parallelism on data scientists and AI practitioners, which is nontrivial and often cumbersome. 

In contrast, ZeRO offers an alternative that requires no model code refactoring. ZeRO is a memory efficient variation of data parallelism \cite{ben2019demystifying,dean2012large} where  model states  are partitioned across all the GPUs, instead of being replicated, and reconstructed using gather based communication collectives on-the-fly during training.   This allows ZeRO to effectively leverage the aggregate GPU memory across machines, at the expense of minimal communication overhead compared to standard data parallel training (2M vs 3M for model size of M) \cite{rajbhandari2020zero}, while still achieving excellent throughput scalability \cite{rajbhandari2021zero}.

\begin{figure}
    \centering
    \includegraphics[width=\columnwidth]{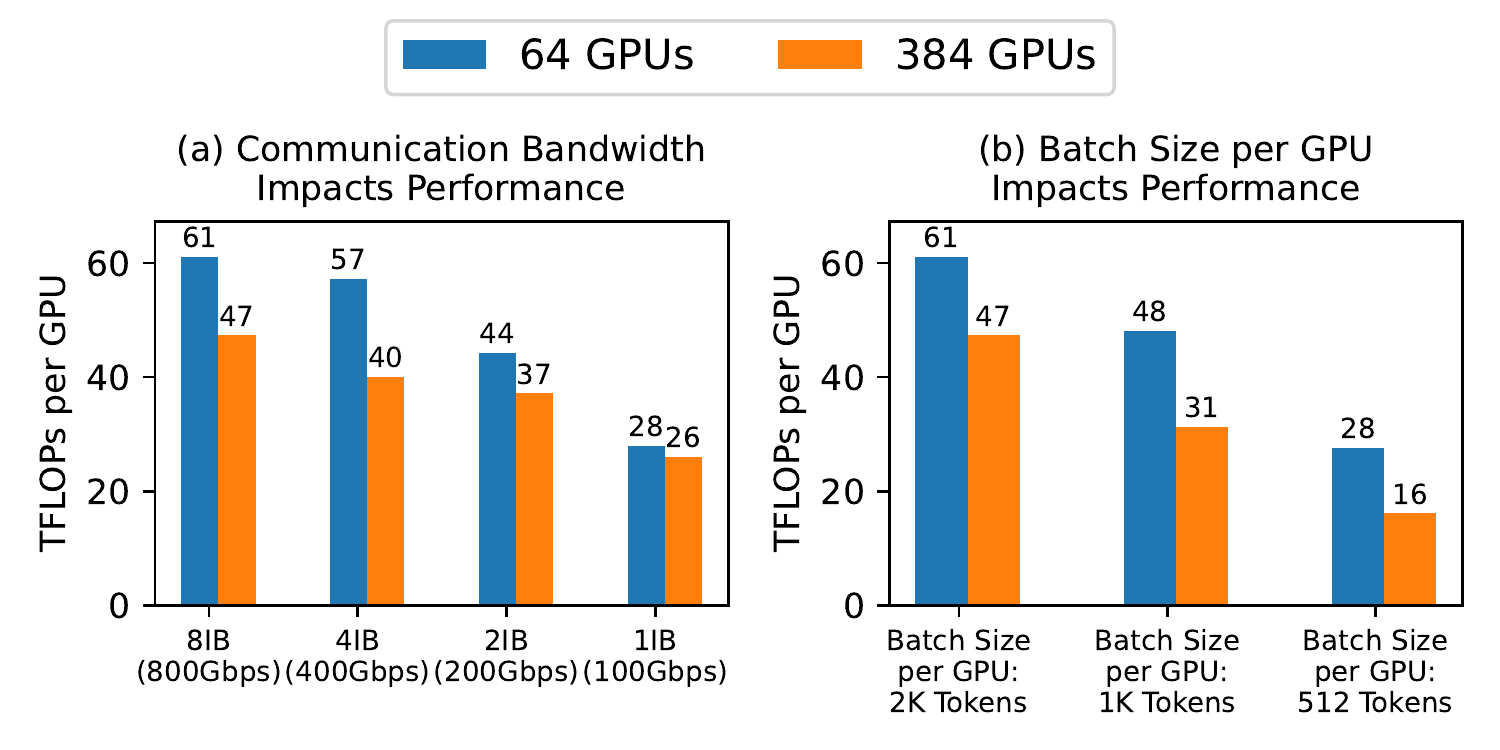}
    \caption{Large scale training throughput are constrained by network bandwidth and batch size per GPU}    \label{fig:intro_plot}
\end{figure}

\subsection{Limitations of ZeRO}
Ease of use of ZeRO combined with its ability to scale efficiently across hundreds to thousands of GPUs, has resulted in its wide adoption. However, there are two critical scenarios where efficiency of ZeRO can be limited due to communication overhead: i) clusters with low-bandwidth, and ii) at very small batch sizes per GPU.
 
On one hand, clusters with low-bandwidth is common in majority of cloud computing environments. Although high performance nodes like DGX boxes~\cite{dgx1,dgx2} are equipped with high-bandwidth NVLink~\cite{nvlink} and NVSwitch~\cite{nvswitch} as intra-node interconnects, cross-node links are often less than 100Gbps ethernet which makes it the communication bottleneck. 
As shown in Figure \ref{fig:intro_plot}(a), the per GPU throughput on low bandwidth clusters is only half of that with high-bandwidth clusters.

On the other hand, even on high-bandwidth clusters, when running on thousands of GPUs, the batch size per GPU is limited by the maximum global batch size that can be used during the training without sacrificing convergence efficiency \cite{DBLP:journals/corr/You-LAMB,ben2019demystifying,keskar2016large}.  In other words, as global batch size cannot be increased indefinitely without slowing down model convergence, training on thousands of GPUs forces the batch size per GPU to be very small, which reduces the compute-to-communication ratio and thus creates a communication bottleneck. 
As shown in Figure \ref{fig:intro_plot}(b), the per GPU throughput is heavily impacted by small batch size per GPU, which is a result of communication bottleneck.

However, rare efforts have been made to optimize end-to-end communication efficiency for ZeRO. There are many previous work on reducing communication overhead in distributed model training, such as 1-bit LAMB \cite{DBLP:journals/corr/abs-2104-06069}, 1-bit Adam \cite{DBLP:journals/corr/abs-2102-02888} and other error compensation compression techniques for gradient averaging \cite{10.5555/3018874.3018875,seide20141,strom2015scalable,alistarh2017qsgd}. However, none of them can work with ZeRO as they all assume model state replication, while model states are partitioned in ZeRO. We start from scratch and provide an end-to-end system for reducing all communication overhead in ZeRO training. 


\subsection{ZeRO++}
In this paper, we present a novel system of communication optimizations collectively called ZeRO++ that offers dramatic communication volume reduction for ZeRO. Below we discuss the main communication overheads in ZeRO, followed by three different communication optimizations in ZeRO++ that address them. 

Assume the model size as $M$. During the forward pass, ZeRO \cite{rajbhandari2020zero} conducts an all-gather operation to collect all the parameters ($M$) needed to train for all model layers. In the backward pass, ZeRO re-collects parameters ($M$) with all-gather first, then each GPU can compute local gradients. After that, ZeRO operates reduce-scatter function to aggregate and redistribute gradients ($M$) across accelerators. In total, ZeRO has a total communication volume of $3M$, spreads evenly across 2 all-gather and 1 reduce-scatter.

To reduce these communication overheads, ZeRO++ has three sets of communication optimizations, targeting each of the above mentioned three communication collectives respectively:  

\textbf{Quantized Weight Communication for ZeRO (qwZ)}  First, in order to reduce parameter communication volume during forward all-gather, we adopt quantization on weights to shrink down each model parameter from FP16 (2 bytes) to INT8 (1 byte) data type before communicating, thus reducing the communication volume by half. However, naively conducting quantization on weights may lose model training accuracy. In order to preserve decent model training precision, we adopt block-based quantization \cite{DBLP:conf/iclr/DettmersLSZ22}, which conducts independent quantization on each subset of model parameters. There is no existing implementation for high performance block-based quantization. Thus, we implement highly optimized quantization CUDA kernels from scratch.  

\textbf{Hierarchical Weight Partition for ZeRO (hpZ)}  Second, to reduce communication overhead of all-gather on weights during backward, we trade GPU memory for communication. More specifically, instead of spreading whole model weights across all the machines, we maintain a full model copy within each machine. At the expense of higher memory overhead, this allows us to replace the expensive cross-machine all-gather on weights with intra-machine all-gather, which is substantially faster due to much higher intra-machine communication bandwidth.  

\textbf{Quantized Gradient Communication for ZeRO (qgZ)} Third, 
reducing communication cost of gradients using reduce-scatter is even more challenging. Directly applying quantization to reduce communication volume is infeasible. The main issue is, even by incorporating block-based quantization to reduce-scatter operation, it will still significantly hurt model training accuracy. The key reason behind is quantization will decrease value precision. And reduction on low-precision values will accumulate and amplify the errors. Therefore, we propose a novel and much more efficient gradient communication paradigm as a general replacement of reduce-scatter collective, where the gradients are compressed using block-based INT4 quantization during the communication to reduce the communication volume, but the full precision is recovered before the reduction operator to preserve training accuracy. We call this $qgZ$, and is designed to i) overcome significant accuracy loss that would result from low-precision reduction if we were to simply implement reduce-scatter in INT4/INT8, and ii) avoid accuracy degradation and significant latency overhead of a long sequence of quantization and dequantization steps needed by a ring \cite{nvidia2017nvidia} or tree \cite{thakur2005optimization,chan2007collective} based reduce-scatter (e.g., left of Figure~\ref{fig:rs-a2a}), even if we did the reductions in full-precision. Furthermore, $qgZ$ leverages the hierarchical nature of modern GPU clusters, where intra-node bandwidth is significantly higher than inter-node, to first reduce gradients within a node before doing cross-node reduction to minimize inter-node communication volume, resulting in 2/4x communication volume reduction (INT8/4) compared to FP16 reduce-scatter. We further reduce end-to-end latency of $qgZ$ by pipelining intra-node and inter-node communication and conducting CUDA kernel fusion.

\textbf {Communication Volume Reduction} By incorporating all three components above, we reduce the cross-node communication volume from $3M$ down to $0.75M$. More specifically, for forward all-gather operation on model weights, by applying INT8 quantization, we reduce the communication size from $M$ to $0.5M$. During backward all-gather on weights, with our secondary copy of model parameters, we reduce the communication size from $M$ to 0. By replacing backward fp16 reduce-scatter on gradients to our novel all-to-all based INT4 reduce-scatter, we reduce cross-node communication from $M$ to $0.25M$. Thus, in total, we reduce $3M$ communication to $0.75M$.

\textbf {Evaluation} We implemented ZeRO++ and performed extensive evaluation demonstrating three key results: i) scalability of GPT-3 like models on up to 384 GPUs achieving over 45\%
of sustained peak throughput, ii) consistent speedup of up to 2.4x over ZeRO \cite{rajbhandari2020zero} baseline across models ranging from 10-138B parameters, and iii) comparing with baseline in 4x higher bandwidth cluster, ZeRO++ achieves similar throughput in low-bandwidth setting. 
In addition, we show the impact of each of the three optimizations in ZeRO++ and how they compose together. Furthermore, we also show the impact of our optimized kernel implementations on end-to-end system throughput. Finally, we conduct convergence evaluation indicating that ZeRO++ has negligible impact on model convergence and maintains similar model training accuracy as ZeRO baseline.

The main contributions of this paper are as follows:
\begin{itemize}
    \item Blocked quantized weights ($qwZ$) reduces communication volume of all-gather of weights by 50\%. 
    \item Hierarchical partitioning of model weights ($hpZ$) completely eliminates inter-node all-gather communication in backward propagation. 
    \item Novel, all-to-all quantized gradient reduction collective ($qgZ$) reduces gradient communication by 75\% comparing with reduce-scatter. 

     \item Optimized Integration of each of the above techniques into existing ZeRO implementation, that enables communication and computation overlapping, and leverages custom high performance CUDA kernels for quantization, dequantization, as well as operator fusion (section \ref{sec:Implementation}). Our implementation translates the 4x communication volume reduction of ZeRO++ into real throughput improvement. 
     \item Extensive experiments shows that i) over 45\% of sustained peak throughput even at small batch sizes, ii) up to 2.4x end-to-end system improvement over ZeRO, and iii) achieving similar throughput in low-bandwidth cluster compared to baseline in high-bandwidth cluster. In addition, we present performance breakdown and analysis of diffrent components of ZeRO++.
      Our end-to-end training 
     shows that ZeRO++ does not affect model convergence.
     \item ZeRO++ is open-sourced and released as part of \url{https://github.com/microsoft/DeepSpeed}
\end{itemize}

\section{Background and Related Work}\label{sec:background}

\subsection{Data, Model and 3D parallelism}
Data parallelism (DP), pipeline parallelism (PP), and tensor parallelism (TP) are three forms of parallelism used to train large models across multi-GPU clusters.  \cite{dean2012large,megatron-2021,narayanan2019pipedream,GPipe} DP is commonly used when model size fits within a single GPU memory. In DP, each GPU holds a full copy of model weights and trains on separate input data.  
MP is orthogonal to DP, and is often used in cases where model size cannot fit into a single GPU's memory. Instead of splitting input data, model parallelism partitions a full model into pieces and assigns each model piece onto a GPU. There are mainly two approaches for model parallelism: i) pipeline parallelism (PP) and ii) tensor parallelism (TP). PP \cite{huang2019gpipe,narayanan2019pipedream,GPipe} splits models vertically, creating sequential stages consisting of a contiguous subset of layers. While there is sequential dependency between stages for an input micro-batch, the stages can be executed in parallel across micro-batches. In contrast, TP \cite{megatron-2021} splits each layer across multiple GPUs, where each GPU works on a different part of the layer for the same input. 


3D parallelism \cite{smith2022using,deepspeed3dblog} refers to combination of DP, PP, and TP, and is capable of achieving excellent throughput and scalability, and has been used to train a wide range of large language models \cite{t-nlg, megatron-2021, gpt-2,gpt-neox-20b}.   
Despite being highly efficient, 3D parallelism is severely limited by the fact that it requires complete rewrite of model and training pipeline to make them compatible with 3D parallelism~\cite{smith2022using}. 






\begin{algorithm}
\small
\caption{ZeRO algorithm}\label{alg:zero}
    \SetKwInOut{KwIn}{Input}
    \SetKwInOut{KwOut}{Output}

    \KwIn{$model$,$worldSize$}
    \KwOut{$model$}

    \While{$model$ not converged} {
       $all\_gather\_Parameters(worldSize)$\;
       $model.forward()$\;
       $partition(worldSize)$\;
       $all\_gather\_Parameters(worldSize)$\;
       $model.backward()$\;
       $partition(worldSize)$\;
       $reduce\_scatter\_Gradients(worldSize)$\;
       $optimizer.step()$\;
    }
    \textbf{Return: } $model$
\end{algorithm}

\subsection{ZeRO Optimizer}
\label{sec:related-zero-opt}
ZeRO is a memory-optimized solution for data parallel training. ZeRO partitions and distributes all model states (i.e., parameters, gradients, optimizer states) among GPUs in use and recollects model states only when the layer needs to be computed. There are three different stages for using ZeRO to optimize on-device memory usage. In ZeRO stage 1 (ZeRO-1), only optimizer states are split and spread across all GPUs in use. ZeRO stage 2 (ZeRO-2) partitions both optimizer states and gradients, where ZeRO stage 3 (ZeRO-3) splits all three components of model states as parameters, gradients, and optimizer states. 

ZeRO-3 is the most memory efficient solution for model training at large scale, but at the cost of more collective communications. Algorithm \ref{alg:zero} illustrates the high-level pseudocode for ZeRO-3. During model training, ZeRO-3 lazy-schedules the fetching of parameters until the computation needs to happen on a particular layer. Before forward propagation, ZeRO launches an all-gather to collect the full model weights and then computes the forward pass (line 2-3) of Algorithm \ref{alg:zero}. Then ZeRO empties the all-gather weights buffer after forward computation completes (line 4). During backward, ZeRO re-collects all model weights again via a second all-gather (line 5) to calculate gradients (line 6). Once gradients are calculated on each GPU, ZeRO empties weights buffer again (line 7) and conducts a reduce-scatter operation to do gradient averaging and re-distribution (line 8). Model states and parameters are updated in optimizer step (line 9).
In a nutshell, to minimize the on-device memory footprint using ZeRO-3, three collective communication operations are issued at each training iteration, which include 2 all-gather on weights and 1 reduce-scatter on gradients. 

\subsection{Communication Reduction Techniques}
\label{sec:related-comm-reduction}
\hspace{3mm}\textbf{Quantization:} Quantization is often used to reduce memory footprint, and data movement volume by using low precision to represent data \cite{dettmers20158,DBLP:conf/iclr/DettmersLSZ22}. However, the loss of information from representing high precision data with lower precision often comes with accuracy degradation. Many related work focus on improving quantization accuracy. The fundamental challenge of quantization accuracy lies in the vast difference in number ranges and granularity between high precision and low precision data (Eg.  FP32/16 vs. INT8). Some related work~\cite{DBLP:conf/icml/ZhaoHDSZ19} propose to filter the outliers in data to mitigate the gap in numerical ranges. Yet their accuracy hinges on the quality of outlier filtering and it brings extra filtering overhead. Dettmers et al.~\cite{DBLP:conf/iclr/DettmersLSZ22} proposes to use block based quantization on optimizer states to improve the quantization accuracy yet it requires changes to the model structure thus limits its usability.

\textbf{Gradient Compression: } 
Starting from 1-bit SGD of error-compensation compression~\cite{seide20141}, gradient compression has been pushed to an extreme direction of using just a single bit. To deal with non-linear gradient-based optimizers like Adam or Lamb, 1-bit quantization algorithms like 1-bit Adam \cite{DBLP:journals/corr/abs-2102-02888} and 1-bit Lamb \cite{DBLP:journals/corr/abs-2104-06069} are proposed, which achieve extreme efficient gradient communication in distributed training. However, 1-bit Adam/LAMB cannot be directly applicable to ZeRO-3. The main reason is 1-bit Adam/Lamb assumes each GPU has the full view of optimizer states (OS) for the model, but ZeRO-3 splits it across all the GPUs in use. Therefore, it is infeasible to directly apply existing gradient compression techniques at ZeRO-3 and we need to design our own.

\textbf{ZeRO Communication Reduction: } 
To reduce expensive cross-node communication, recent optimization on ZeRO-3, such as MiCS \cite{zhang2022mics}, trades on-device memory for communication. In MiCS, the GPU cluster is divided into sub-groups, and model states are partitioned within a sub-group but replicated across sub-groups. By keeping the sub-group size small, MiCS can either leverage high bandwidth intra-node interconnect, or use hierarchical communication to lower the communication volume.  $hpZ$ in ZeRO++ adopts a similar approach of trading memory for less communication. The key difference is that $hpZ$ only do secondary partition on weights, while keeping all other model states partitioned across all GPUs. This allows hpZ to achieve significant communication reduction without the massive memory overhead of MiCS.

\section{Design}
In this section, we elaborate on the design of our three key optimizations in ZeRO++ introduced in Section~\ref{sec:intro} for reducing the communication overhead of  ZeRO: i) Quantized Weight Communication for ZeRO ($qwZ$), ii) Hierarchical Partitioning for ZeRO ($hpZ$), and iii) Quantized Gradient communication for ZeRO ($qgZ$). After that, we discuss the end-to-end impact of these optimizations to reduce to total communication volume of ZeRO. 

\subsection{Quantized Weight Communication for ZeRO ($qwZ$) }

\begin{figure}
    \centering
    \includegraphics[width=\columnwidth]{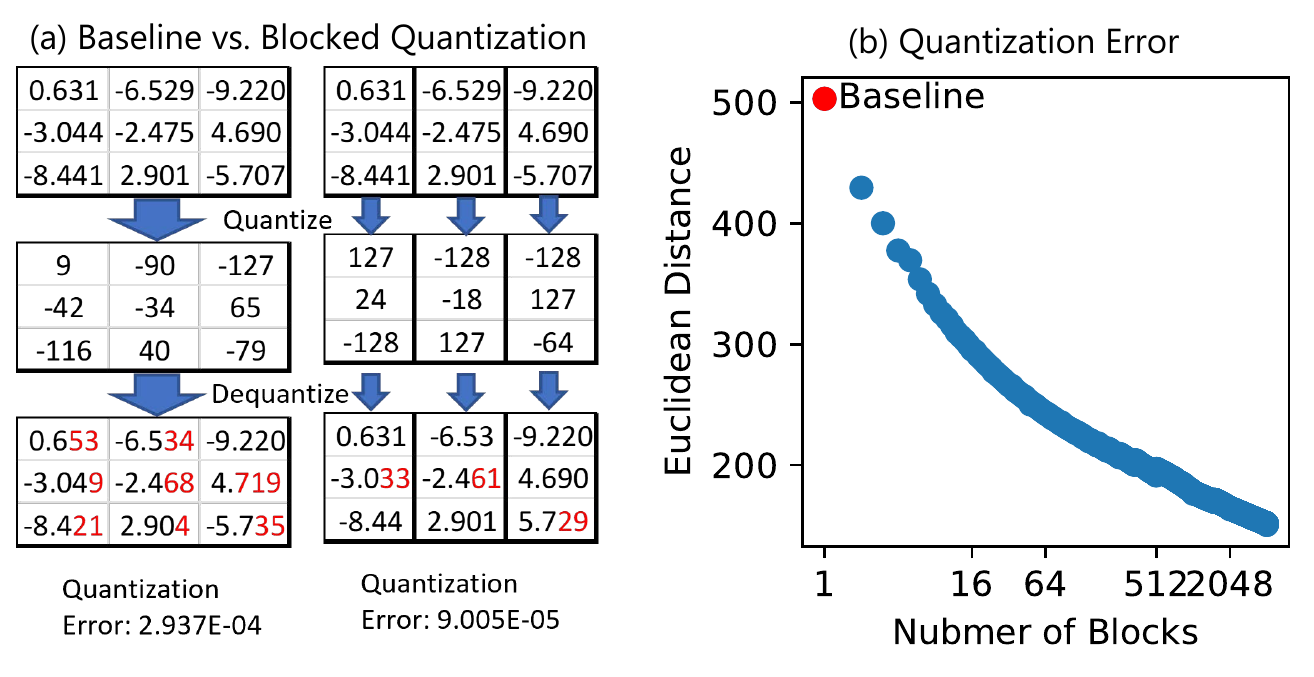}
    \caption{Illustration \& example of block based quantization vs. baseline}    \label{fig:methodology_qw}
\end{figure}

As discussed in Section~\ref{sec:related-zero-opt}, ZeRO partitions the model weights across all the ranks (i.e., GPUs) and fetches the FP16 weights layer-by-layer right before they are needed in computation via all-gather for the forward and backward of each training iteration. To reduce the communication overhead of forward all-gather on weights, $qwZ$, quantizes FP16 weights to INT8 right during the all-gather, and dequantizes them back to FP16 on the receiver side, and then conducts layer computation. 

While this reduces the communication volume of the all-gather by 2x, doing so naively results in two major issues: i) the lowering of precision results in significant accuracy degradation during training as discussed in ~\ref{sec:related-comm-reduction}
, and ii) the quantization and dequantization overhead negates any throughput gain from communication volume reduction. We discuss the optimized implementation of $qwZ$ to minimize the quantization and dequantization overhead in Section~\ref{sec:Implementation}. Here, we primarily focus on design choices to mitigate accuracy degradation.


$qwZ$ uses blocked based quantization to improve the quantization accuracy. As illustrated in Figure \ref{fig:methodology_qw}, each weight tensor is divided into smaller chunks, and converted into INT8 by symmetric quantization, using an independent quantization scaling coefficient. By keeping the quantization granularity small, we significantly mitigate the gap in number ranges and granularity. 

We show an example of the quantization error of performing block based quantization vs. the non-blocked quantization baseline in Figure \ref{fig:methodology_qw}(a). Fig.~\ref{fig:methodology_qw}(b) shows a case study of weights quantization on BERT model, where block based quantization reduces the quantization error by 3x. More in-depth convergence evaluations are shown in Sec. \ref{sec:eval}.




\subsection{Hierarchical Partitioning for ZeRO ($hpZ$)}

 \begin{figure}
\centering
\includegraphics[width=0.49\textwidth]{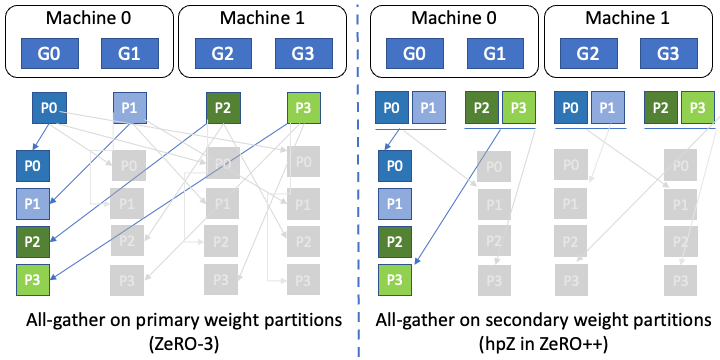}
\caption{\label{fig:zero_hpZeRO} hpZ removes cross node traffic in backward all-gather by holding secondary weight partitions in on-device memory.}
\end{figure}

ZeRO-3 partitions all its model states across all its ranks, resulting in communication collectives that span all the GPUs. With $hpZ$,  we notice that it is possible to have different partitioning for different model states, limiting the communication collectives to a subset of the GPUs. Given that on modern GPU clusters, intra-node communication bandwidth is significantly higher than inter-node communication bandwidth, this presents opportunities to reduce the inter-node communication.

More specifically, in $hpZ$, we eliminate the inter-node all-gather during the backward pass by holding secondary FP16 weights partition within each node. We do this by creating a hierarchical partitioning strategy consisting of two partitions: first, all model states are partitioned globally across all devices as in ZeRO-3, which we call primary partition. Second, a secondary copy of FP16 parameters is partitioned at the sub-global level (e,.g., compute node, see figure~\ref{fig:zero_hpZeRO}), which we call secondary partition. This secondary copy of FP16 parameters is replicated across multiple secondary partitions.

Consider a 64-node cluster, each node with 8 GPUs. Model weights are partitioned in two stages: i) across all 512 GPUs that we call primary partition, and ii) the same weights are also partitioned within a compute node across 8 GPUs, that we call secondary partition. In this example, for the secondary partition, each compute node in the cluster holds a full replica of FP16 weights partitioned among the 8 GPUs within the node, and there are 64 of such replicas in total.


\subsubsection{A training iteration with hpZ} 
During the forward pass of a training iteration, we all-gather weights based on the primary partition across all GPUs. However, once the weights are consumed during the forward pass, they are partitioned based on the secondary partition. Given the temporal consistency of model parameters between forward and backward passes, when the weights are needed again during the backward pass, we all-gather weights based on this secondary group. Note that when the secondary partitioning is set to be a compute node, this avoids any inter-node communication for this all-gather. Finally, at the end of the iteration, during the optimizer step, all the model states, as well as the  primary copy of the fp16 parameter are updated based on the primary partition. hpZ makes two changes to baseline ZeRO pseudocode in Algorithm \ref{alg:zero}: i) in line 4, parameter partitioning is based on \emph{secondary group size}, ii) parameter all-gather preceding backward pass in line 5 is also based on \emph{secondary group size}.  

Our design of $hpZ$ is flexible to support any \emph{secondary group size}. The group size controls how many ranks (i.e., GPUs) are in the secondary partition. It is also a measure of memory-communication trade-off of $hpZ$. Simply put, by default, $hpZ$ secondary partition is node-based (recall intra-node bandwidth is multiple factors of inter-node bandwidth for current and future hardware configurations) but can be extended to support multiple compute nodes as needed. 




\begin{figure}[t]
\centering
\includegraphics[width=0.49\textwidth]{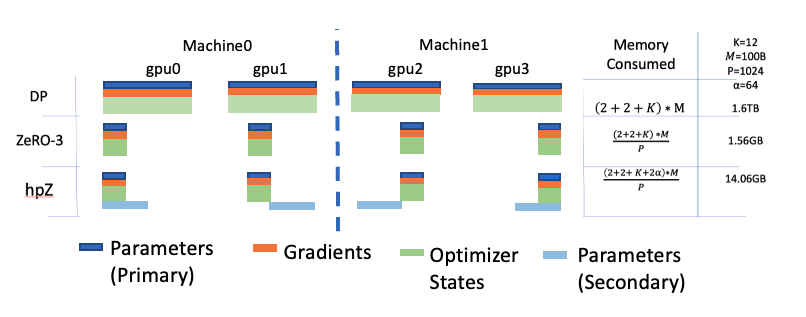}
\caption{\label{fig:hpz_memory_usage}Per-device memory consumption analysis of standard data parallel (DP), ZeRO stage 3 (ZeRO-3) and proposed hierarchical partitioning of ZeRO parameters ($hpZ$). $K$ denotes the memory multiplier of optimizer states, $M$ represents the number of trainable parameters, $P$ is the data parallel group size or world size, and $\alpha$ is the number of secondary groups or ratio of world size to the number of ranks in the secondary group. A typical real world scenario example is provided in the last column. We assume a model size of 100B trained on 1024 V100 GPU DGX cluster (64 compute nodes, 16 GPUs per node).}
\end{figure}

\subsubsection{Memory Usage Analysis}
By design, $hpZ$ trades memory for communication efficiency. It is important to analyze this tradeoff. Recall that standard data parallel DNN (DP) replicates model parameters across data parallel ranks, ZeRO-3 on the other hand partitions parameter across data parallel ranks. A midway approach is model parameters partitioned across a subset of devices as long as model parameters fit. 

Figure \ref{fig:hpz_memory_usage} provides a concrete memory usage estimate of a typical large language model of size of 100B parameters, with primary group size of 1024 GPUs and secondary group size of 16 GPUs (e.g., DGX-2 V100 node). As shown in Figure \ref{fig:hpz_memory_usage}, with our proposed method, $hpZ$ consumes $8.9x$ more memory than ZeRO-3, our approach is still $114x$ less memory requirement than standard DP. 
This marginal increase in memory usage is compensated for by efficient intra-node communication schedule. By eliminating or reducing inter-node communication for backward pass, $hpZ$ reduces the end-to-end communication of ZeRO by $1.5x$, while still supporting model training with hundreds of billions of parameters. 

\subsection{Quantized Gradients Communication for ZeRO ($qgZ$)}
\label{sec:design-qgz}





In this section, we propose a novel quantized reduce-scatter algorithm called qgZ based on all-to-all collectives that enables a 4x communication volume reduction of gradient reduce-scatter by replacing FP16 with INT4 quantized data, while overcoming precision loss challenges described in Section~\ref{sec:intro}, as well as numerous system challenges that we will outline in this section. 

qgZ leverages all-to-all collectives to implement quantized reduce-scatter which includes three major components: 1) all-to-all-based implementation of quantized gradient reduce-scatter, 2) reducing communication volume with hierarchical collectives, 3) tensor slice reordering for correct gradient placement. We talk about each of them step-by-step.



\subsubsection{ All-to-all based implementation}
\label{sec:design-qgz-1hop}

\begin{figure}[t]
\centering
\includegraphics[width=0.5\textwidth]{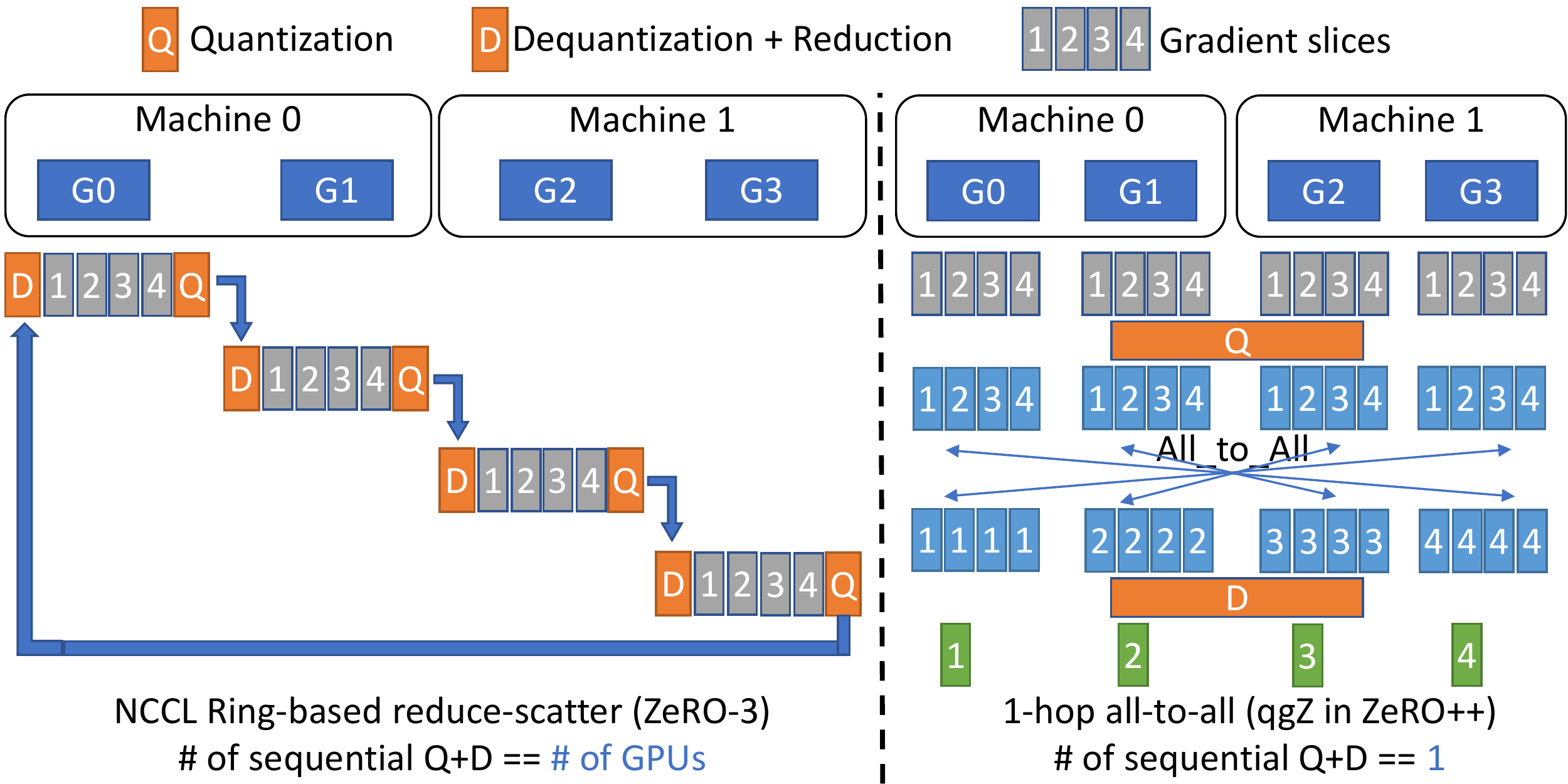}
\caption{\label{fig:rs-a2a} Comparison between ZeRO-3 ring-based reduce-scatter and qgZ 1-hop all-to-all.}
\end{figure}

A naive approach towards quantized reduce-scatter, while avoiding precision loss due to reduction is to apply quantization and dequantization to a ring-based reduce-scatter directly as shown on the left of Figure~\ref{fig:rs-a2a}. We can inject quantization and dequantization on each GPU. Once a GPU receives gradients from its predecessor, we dequantize it to recover full precision and conduct a local reduction. Next we can quantize local reduction output and pass quantized data to its successor. To finish the whole reduce-scatter, the number of sequential quantization and dequantization kernels is equal to the number of GPUs (i.e., n) in use. 

Thus, applying quantization and dequantization on existing ring based reduce-scatter collective will lead to high communication latency and low value precision due to multiple sequential quantization and dequantization steps. Although recent tree-based collective like Blink\cite{DBLP:conf/mlsys/WangVPTDS20} could reduce the number of sequential kernels from n to log(n), the long latency and low precision issue is not completely resolved.

To overcome this, we completely abandon existing ring-based reduce-scatter approach and incorporate 1-hop all-to-all collective for our gradient communication. As shown on the right of Figure~\ref{fig:rs-a2a}, we first apply quantization on a given tensor, then we conduct all-to-all communication among all the GPUs. After all-to-all, we apply another dequantization to recover the data precision and then reduce on high-precision values to get the final gradient reduction output. By replacing ring-based solution with our all-to-all 
collective, we reduce the number of sequential quantization+dequantization kernel from the number of GPUs to 1. Thus, we solve the long latency and low precision issues when applying quantization in reduce-scatter for supercomputing scenarios like DGX boxes connected in fat-tree topology. 

\begin{figure}[t]
\centering
\includegraphics[width=0.5\textwidth]{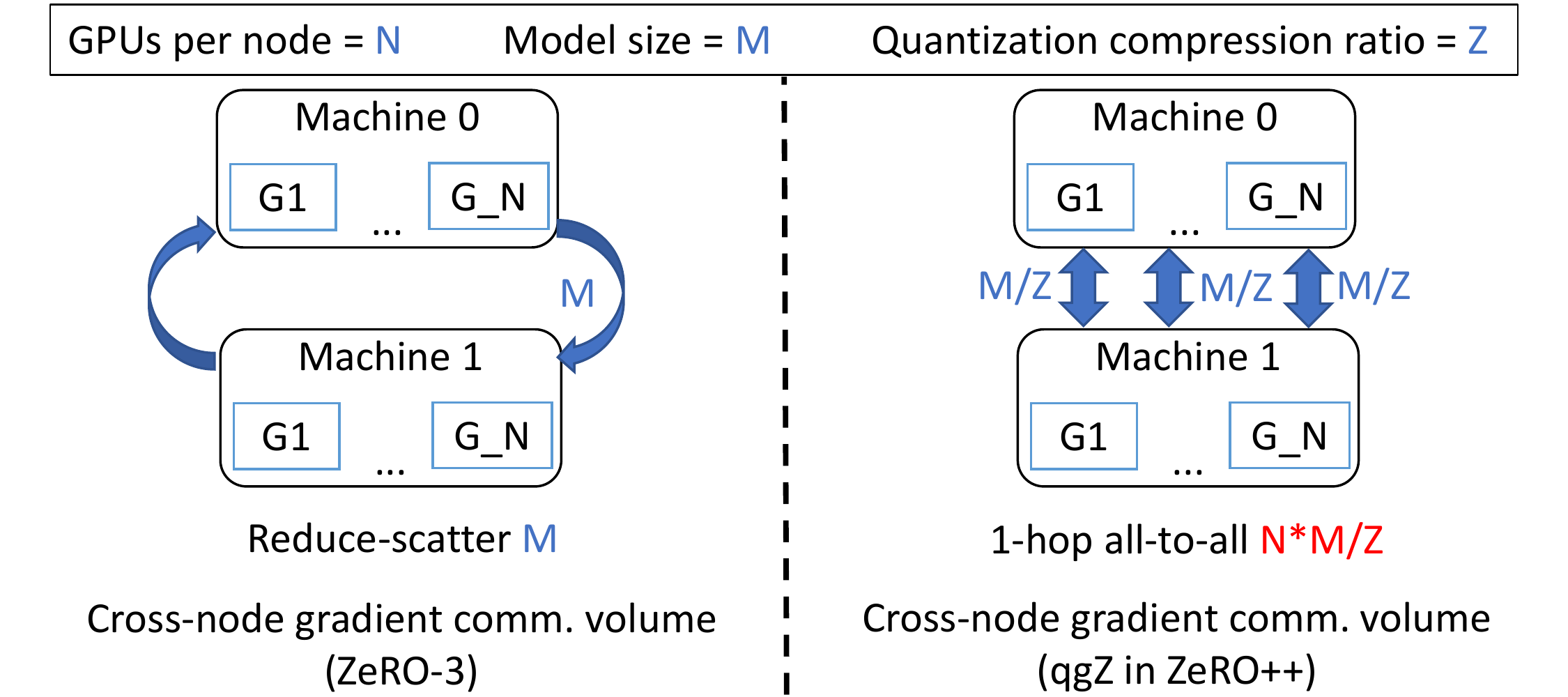}
\caption{\label{fig:volume-compare} Communication volume comparison between ZeRO-3 reduce-scatter and qgZ 1-hop all-to-all.}
\end{figure}

\subsubsection{Reducing inter-node communication volume}
\label{sec:design-qgz-2hop}

Although replacing reduce-scatter with all-to-all achieves single-shot quantization and dequantization, it introduces a new problem; the inter-node communication volume increases instead of decreasing despite the quantization of data. We elaborate on this in Figure \ref{fig:volume-compare}.


Here we assume model size of $M$, GPU per node is $N$, gradient compression ratio as $Z$. Reduce-scatter, reduces the data during transmission over the ring, thus the total amount of data for cross-node communication is M. However, when using our 1-hop all-to-all approach, even though the data are compressed before communication (i.e., $M/Z$), each GPU needs to send out $M/Z$ amount of data to GPUs on the other nodes. Therefore, each machine will generate $N*M/Z$ amount of cross-node communication data, which is much bigger than reduce-scatter communication volume.


\begin{figure}[t]
\centering
\includegraphics[width=0.5\textwidth]{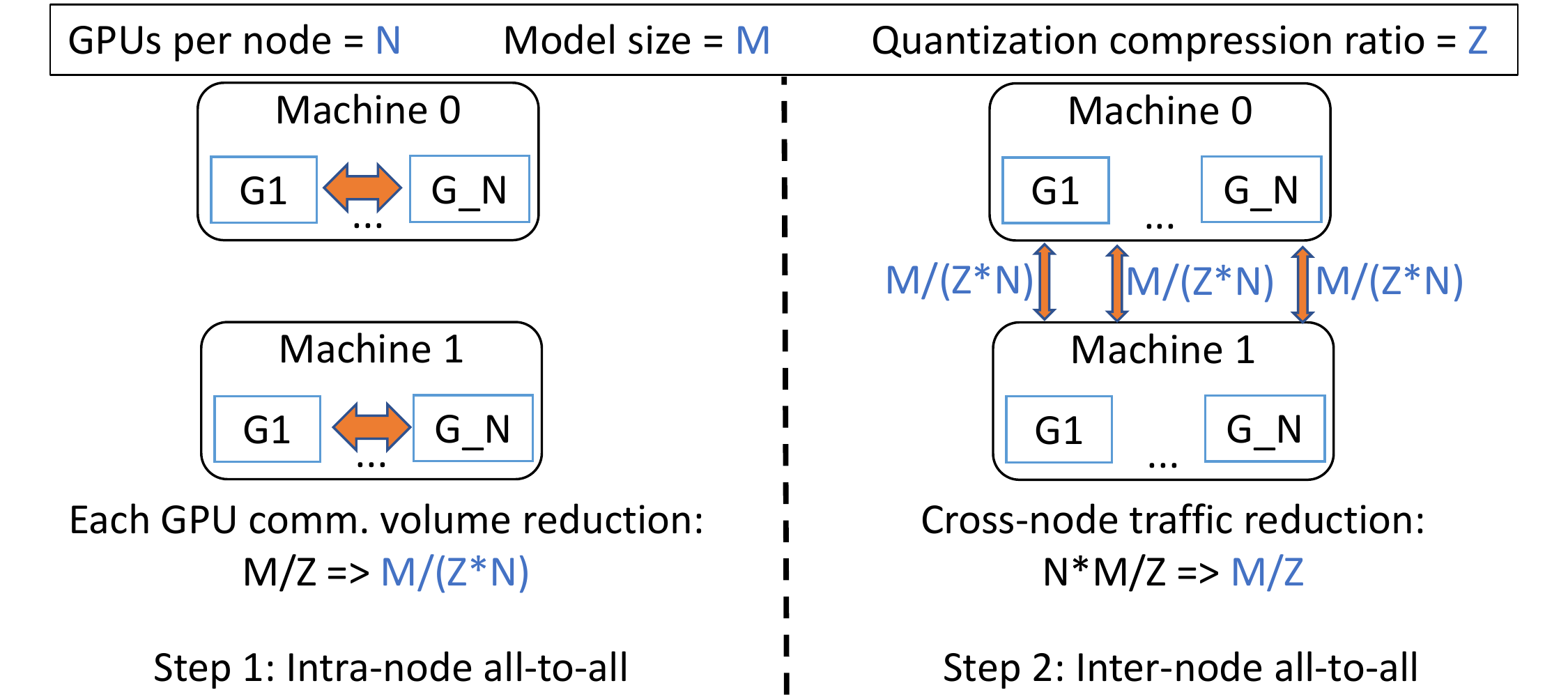}
\caption{\label{fig:qg-hierach-all2all} qgZ apply hierarchy all-to-all to reduce cross node traffic.}
\end{figure}

To address this, we do a hierarchical 2-hop all-to-all instead of 1-hop: a) first intra-node all-to-all and b) followed by inter-node all-to-all, which is shown as Figure~\ref{fig:qg-hierach-all2all}. First, with high-bandwidth links among GPUs inside a machine, we conduct intra-node all-to-all on quantized data, then dequantize data and reduce on dequantized data. After intra-node quantization, all-to-all, dequantization, and reduction, we reduce the data size per GPU from $M/Z$ to $M/(Z*N)$. After intra-node all-to-all is completed, we conduct the inter-node all-to-all communication, which is similar to 1-hop all-to-all we described above. Given that now each GPU only needs to send out $M/(Z*N)$ data, the communication volume per machine is now $M/(Z*N) * N = M/Z$. By adopting this hierarchical all-to-all communication as 2-hop approach, we resolve the communication volume blow-up issue in our 1-hop scheme perfectly. Note that even though the total communication volume is doubled (one intra-node, the other inter-node), intra-node communication introduces negligible overhead given NVLink/NVswitch high bandwidth, and cross-node traffic has been significantly reduced, which is the major bottleneck in gradient communication. 

\subsubsection{Tensor slice reordering for correct data placement}
\label{sec:design-qgz-reorder}
\begin{figure}[t]
\centering
\includegraphics[width=0.5\textwidth]{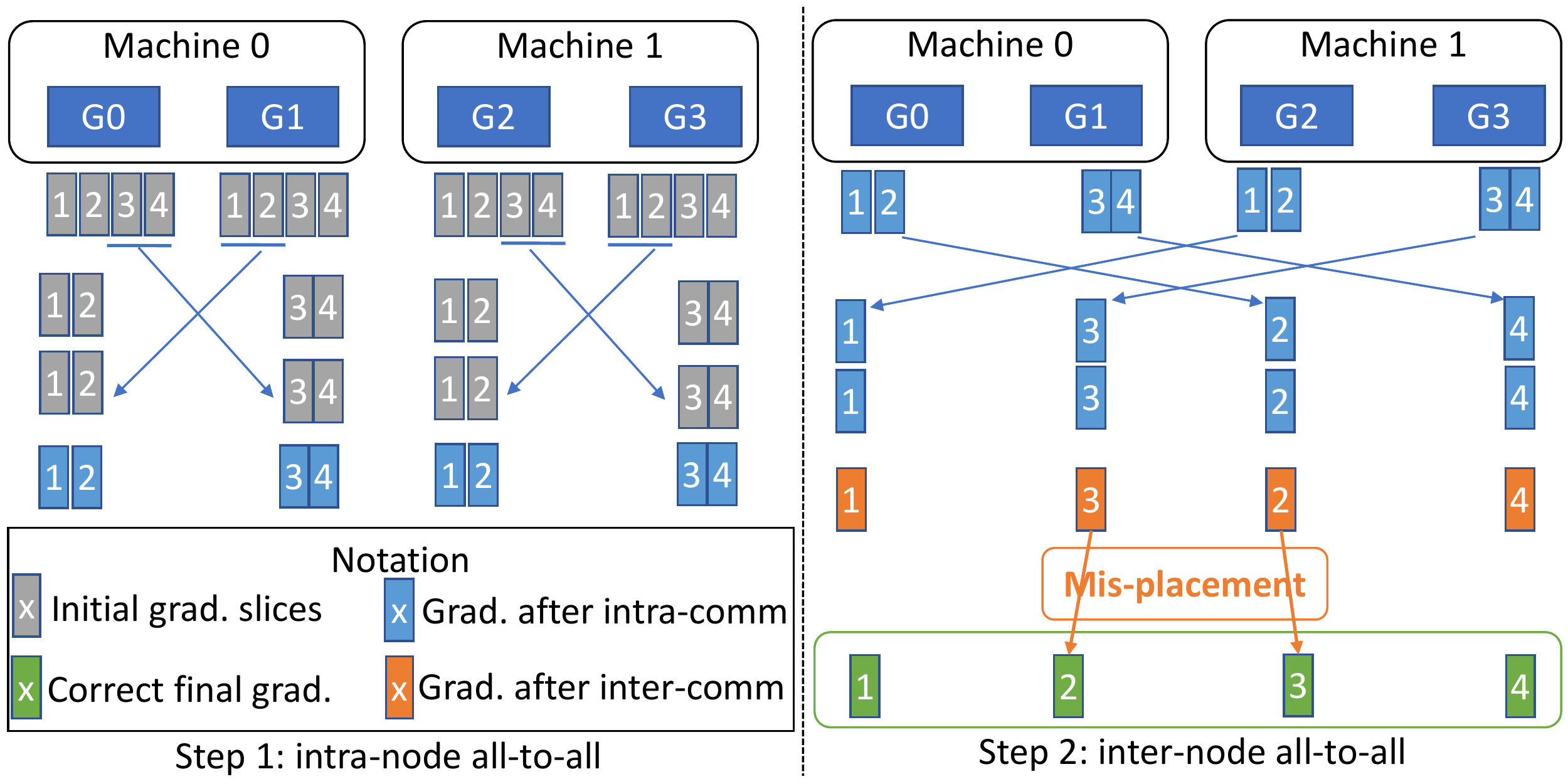}
\caption{\label{fig:data-misplacement} Gradient partition misplacement when applying hierarchical all-to-all in qgZ.}
\end{figure}

With the 2-hop all-to-all, the inter-node communication volume is as expected, however, this introduces a gradient misplacement issue. We describe this issue using a 2x2 example, where we have 2 machines and each machine has 2 GPUs. As shown in Figure~\ref{fig:data-misplacement}, the correct final gradient placement is shown as green boxes in the figure, where GPU 0 holds final gradient partition 1, GPU 1 holds gradient partition 2, so on and so forth. 

Our 2-step all-to-all communication works as follows, first we divide all gradients on each GPU into 4 chunks, then conduct our intra-node all-to-all. After intra-node all-to-all finishes, GPU0 (i.e., G0) holds partial aggregated gradient partition 1,2 whereas G1 holds gradient partition 3,4. Same thing happens on G2 and G3. Since G1 does not have gradient partition 2 (which is supposed to be held by G1) while G2 does not have gradient partition 3, after inter-node all-to-all, there is gradient misplacement issue on both G1 and G2. 

\begin{figure}[t]
\centering
\includegraphics[width=0.5\textwidth]{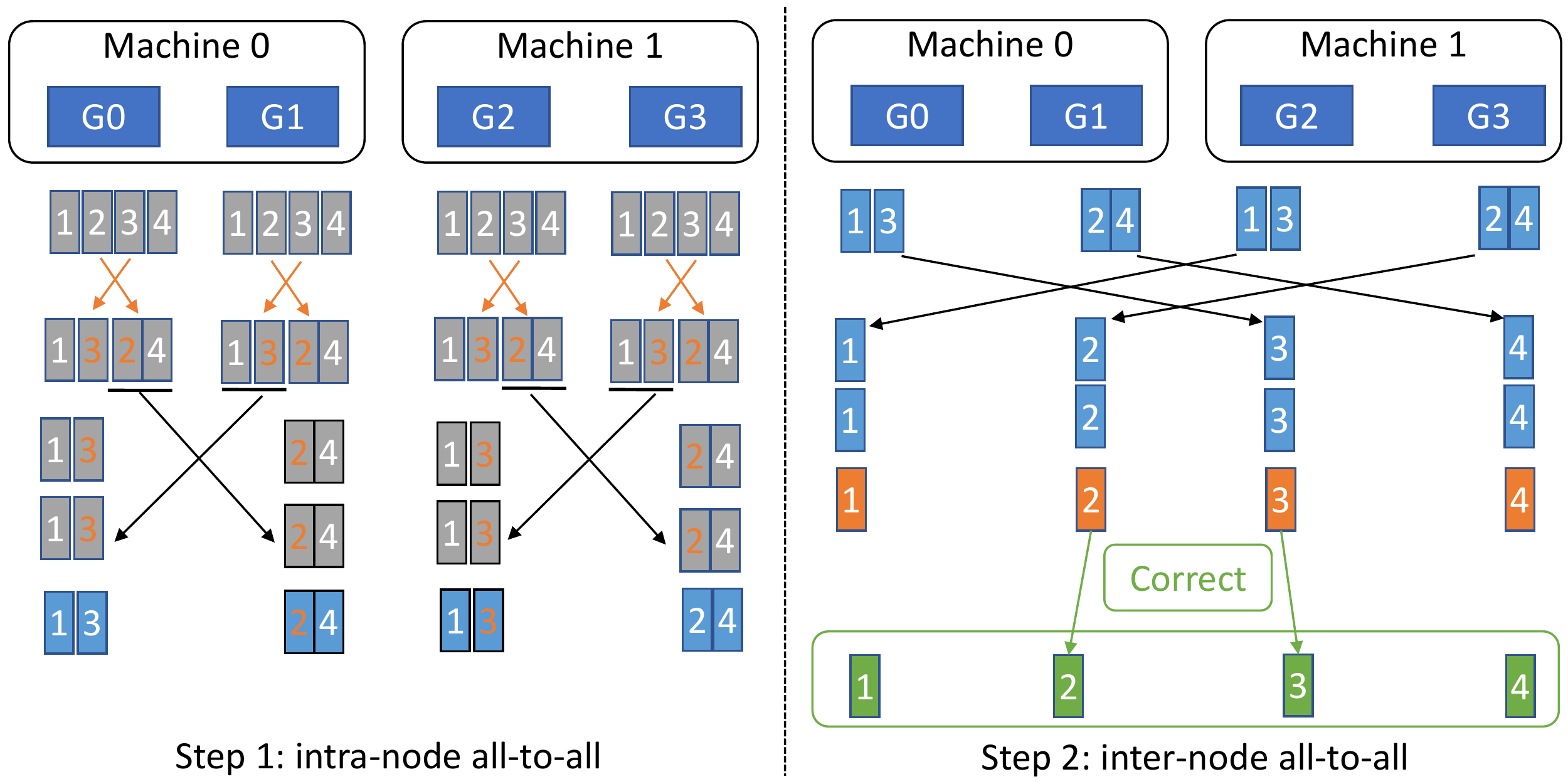}
\caption{\label{fig:tensor-reorder} Tensor slices reordering to correct gradient misplacement in qgZ.}
\end{figure}

We address this with tensor slice reordering. As shown in Figure~\ref{fig:tensor-reorder}, before intra-node all-to-all begin, we first swap the tensor slice order of slice 2 and 3, which is shown as orange arrows. Then after intra-node all-to-all is completed, G1 now has gradient 2 while G2 has gradient 3. Therefore, after the inter-node all-to-all, all GPUs get the correct gradient placement. Mathematically, given X GPUs per node and Y nodes in total, each GPU will hold X*Y gradient slices initially. Our tensor slice reordering works as follows:
\begin{equation}
before: [0, 1, 2, 3, 4, ... YX-3, YX-2, YX-1]
\label{eq:swizzle-1}
\end{equation}
\begin{equation}
after: [0, X, 2X, ... (Y-1)X, 1, X+1, (Y-1)X+1, ... YX-1]
\label{eq:swizzle-2}
\end{equation}


Based on Equation \ref{eq:swizzle-1} and \ref{eq:swizzle-2}, we can map each original tensor slice position (i.e., Equation \ref{eq:swizzle-1}) to new tensor slice position (i.e., Equation \ref{eq:swizzle-2}) on each GPU to correct final gradient misplacement issue. 

In summary, by solving above three challenges step-by-step, we design a novel gradient communication and reduction protocol, which can be a more communication efficient and generalized replacement of reduce-scatter collective. We discuss some of the optimization and implementation details for our approach in Sec.~\ref{sec:Implementation}.


\subsection{ZeRO++ Communication Volume Analysis}

Table \ref{tb:comm-analysis} illustrates theoretical communication volume comparison between ZeRO-3 and ZeRO++. We assume the model size of $M$. As described in Section 2, during ZeRO-3 there are 3 collective calls: all-gather on weights in forward pass, then all-gather on weights in backward pass and last is reduce-scatter on gradients in the backward. And each collective communicates $M$ volume of data. 


With ZeRO-3, in total we need to communicate 3M data per each training iteration. Given that intra-node communication is fast with NVLink and NVSwitch, we ignore intra-node communication and focus on cross-node traffic only. For all-gather in the forward pass, by incorporating our quantized weights communication, we reduce communication volume from M to 0.5M. During the all-gather in the backward pass, by holding secondary weights partition within each node, we completely removed cross-node traffic. For reduce-scatter in the backward pass, by replacing reduce-scatter with our novel quantized gradient communication protocol, we reduce cross-node traffic from M to 0.25M. Therefore, compared with ZeRO-3, ZeRO++ reduces communication volume from 3M down to 0.75M for each training iteration. 

\begin{table}[t]
\centering
\begin{tabular}{ |c|c|c|c| } 
 \hline
Comm. & forward & backward & backward \\

Volume & all-gather & all-gather & reduce-scatter \\
\hline
ZeRO-3 & M & M & M \\ 
\hline
ZeRO++ & 0.5M & 0 & 0.25M \\ 
\hline
\end{tabular}
\caption{\label{tb:comm-analysis} Communication volume comparison between ZeRO-3 and ZeRO++.}
\end{table}

\section{Optimized Implementation}\label{sec:Implementation}



In this section, we discuss two key optimizations that enable ZeRO++ to fully realize the potential of
4x communication volume reduction to improve throughput without getting limited by implementation overheads: i) overlapping different communication and compute streams, when doing so enables better resource utilization, and ii) optimized CUDA kernels for quantization, dequantization, and tensor slice reordering operators, and kernel fusion across these operators  when appropriate to minimize the memory traffic overhead. Below we discuss the two lines of optimization in detail.

\subsection{Overlap Compute and Communication}

To reduce end-to-end communication time, we overlap quantization computation with communication for all-gathering of weights in both forward and backward passes. For the hierarchical all-to-all based reduce-scatter implementation of gradients, we overlap the intra-node communication with inter-node communication.



\subsubsection{Communication-computation overlapping on weights}
For all-gather on weights, we enable communication-computation overlap using two key features : i) we track the execution order of model layers to get the sequence they will be fetched.
ii) we guarantee asynchronous quantization execution.
Specifically,
the call to the quantization kernel is non-blocking and
we further avoid operations that involve explicit/implicit CUDA synchronization (e.g. tensor concatenation), making the quantization a non-blocking operation that can be launched asynchronously.

With this two features, as ZeRO fetch parameters for each layer, the communication of the current layer and the quantization of the next layer can be launched at the same time on different CUDA streams. When the quantized data are needed for the next layer, ZeRO++ synchronizes the quantization stream to make sure the quantized data are ready. This approach hides the quantization cost of the next layer under the communication time span of the current layer which hides the quantization overhead.

\begin{figure}[t]
\centering
\includegraphics[width=0.48\textwidth]{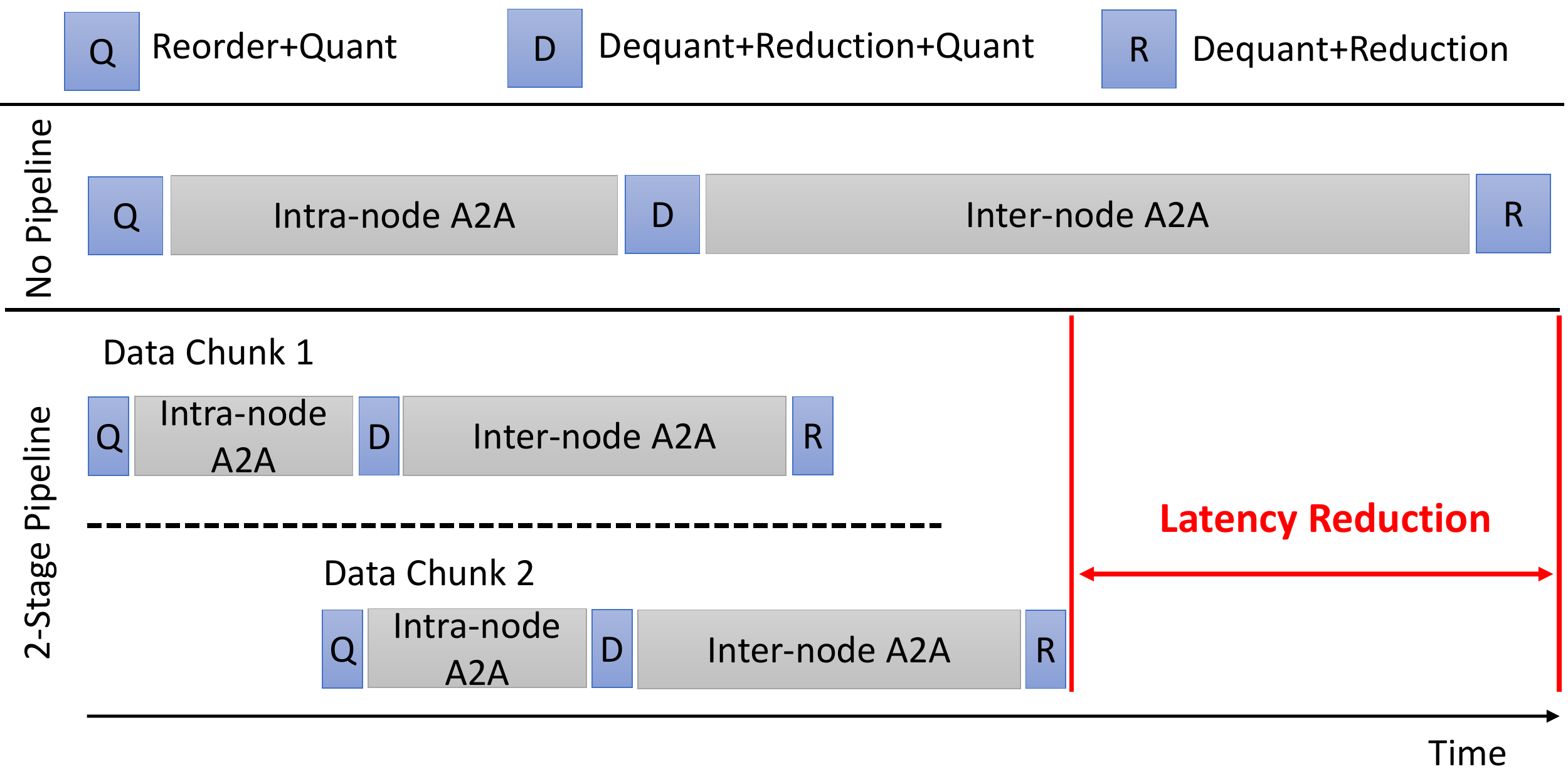}
\caption{\label{fig:qg-pipeline} Pipelining and overlapping intra-node communication with inter-node communication in $qgZ$.}
\end{figure}

\subsubsection{Hierarchical Collectives for Gradient Communication}

As discussed in Sec.~\ref{sec:design-qgz-2hop}, our all-to-all based gradient communication is broken into two stages: first intra-node communication followed by inter-node communication. The inter-node communication depends on the results of the intra-node communication, therefore, with a naive implementation, inter-nodes links are idle during intra-node communication and vice versa. To reduce latency by leveraging both inter-node and intra-node links in parallel, we chunk our input gradient tensor and pipeline transfer between intra-node communication and inter-node communication. As shown in Figure~\ref{fig:qg-pipeline}, compared with "no pipeline" case on the top, simply adopting a "2-stage pipeline" transfer achieves the amount of end-to-end latency reduction shown as the red arrow-line in Figure~\ref{fig:qg-pipeline}. By overlapping intra-node and inter-node communication, the end-to-end latency of gradient communication is significantly reduced. 

\begin{algorithm}
    \SetKwInOut{KwIn}{Input}
    \SetKwInOut{KwOut}{Output}
    \SetKwInOut{KwConst}{Constants}

    \KwConst{$stages$, $nodeSize$, $nodes$}
    \KwIn{$partitionID$}
    \KwOut{$mappedPartitionID$}

    $totalDevices \gets nodeSize * nodes$\;
    $stageID \gets partitionID \mathbin{\%} totalDevices$\;
    $chunkID \gets \frac{partitionID}{totalDevices}$\;

    $pipelineOffset \gets stageID * totalDevices$\;

    $chunkOffset \gets \frac{stageID}{nodeSize}$\;
    $chunkBase \gets (chunkID \mathbin{\%} nodeSize) * nodes$\;

    \textbf{Return: } $pipelineOffset + chunkBase + chunkOffset$\;

    \caption{Generalized tensor slice reordering ($qgZ$)}
    \label{algo:swizzle-pipe}
\end{algorithm}
Doing this pipeline correctly has implications on our tensor slice reordering process. The more pipeline stages we have, the more fine-grained tensor slices are needed for reordering. Therefore, we also propose a generalized tensor slices reordering scheme as algorithm~\ref{algo:swizzle-pipe}, which covers both w/ and w/o pipelining data transfer cases. Here stages refer to the number of pipeline stages we have, nodeSize is the number of GPUs per node and nodes is the number of nodes.


Next, we discuss how we optimize our CUDA kernels to further reduce all quantization related overhead.

\subsection{CUDA Kernels}


As existing quantization implementations are unable to capture the combination of data mapping and high throughput necessary to minimize kernel overhead, we implement and optimize custom CUDA kernels to implement these primitives. In particular, these kernels aim to (1) saturate device memory bandwidth and (2) minimize the total traffic via fusion.

\textbf{Maximizing Bandwidth Utilization:} A core quantization and dequantization library of composable operators was developed as the foundation for ZeRO++. The core primitives leverage efficient vectorized memory accesses at the maximum granularity a given GPU architecture supports. In order to satisfy the alignment requirements these instructions have, model state is partitioned such that quantization granularities will be 16B aligned. Additionally, we leverage instruction level parallelism to overlap multiple memory transactions with each other. In practice, the combination of vectorized accesses and instruction level parallelism enables the quantization library to achieve full GPU memory bandwidth utilization. 


\textbf{Minimizing Total Traffic:} Multiple techniques are used to reduce the total memory traffic for quantization kernels. First, the size of each quantization block is tuned so as to express sufficient parallelism to schedule across a GPU's streaming multiprocessors and cache values not quantized yet in the register file while calculating the quantization scale and offset for the block. 
Second, we fuse tensor reshaping and quantization into the same kernel to avoid redundantly loading data from global memory. For example, the tensor slice reordering 
(i.e., orange arrow-lines in Figure ~\ref{fig:tensor-reorder}) is realized within a fused quantization and remapping kernel.
This fused kernel achieves the same level of performance as a single quantization kernel working with contiguous data. 
Finally, we fuse sequential dequantization, reduction, and quantization operations into single kernel implementation, which reduces total memory traffic by ~9x in $qgZ$.

\section{Evaluation}\label{sec:eval}

In this section, we perform three sets of evaluations for ZeRO++. First, we perform end-to-end evaluations showing : i) scalability evaluation on up to 384 GPUs , ii) speedup over state-of-the-art (SOTA) baseline across models ranging from 10-138B parameters, and iii) throughput comparisons for cluster setting with varied cross-node bandwidth. Second, we perform throughput analysis and breakdown, evaluating the impact of different components of ZeRO++, as well as the impacts of our kernel optimizations on end-to-end throughput. Finally, we show convergence evaluation indicating that ZeRO++ doesn't harm model convergence and maintains similar model training accuracy.

\subsection{Methodology}

\textbf{Hardware: } 24 NVIDIA DGX-2 nodes where each with 16 V100 SXM3 32 GB GPUs~\cite{dgx2}. The nodes are connected by InfiniBand (IB) with NVIDIA SHARP support~\cite{ib-sharp}, achieving total inter-node bandwidth of over 800 Gbps. To evaluate ZeRO++ in clusters under different network environments, we show the performance of ZeRO++ running with different cross-node bandwidth by enabling from 1 to 8 IB connections (i.e., 100 Gbps to 800 Gbps).

\noindent\textbf{Baseline:} We use ZeRO-3 as the baseline given its ease-to-use for training giant models at large scale. To evaluate the performance of our optimized kernels, we also implemented ZeRO++ with PyTorch quantization\cite{pytorch-quant} and non-fused kernels as baselines for our ablation study.

\noindent\textbf{Model Configurations:} We use GPT-style transformer models for evaluation. Given Megatron-Turing-NLG~\cite{smith2022using} training 530B model on 2K GPUs using 2K tokens per GPU (i.e., micro batch size), we evaluate ZeRO++ with the same 2k tokens per GPU setting. We also evaluate on 1K tokens per GPU to test ZeRO++ with more extreme scale scenario. The number of layers and hidden sizes are adjusted to have models of different sizes.
Please refer to the appendix and our open-sourced evaluation scripts for hyperparameters and other training details.

\begin{figure}
    \centering
    \includegraphics[width=\columnwidth]{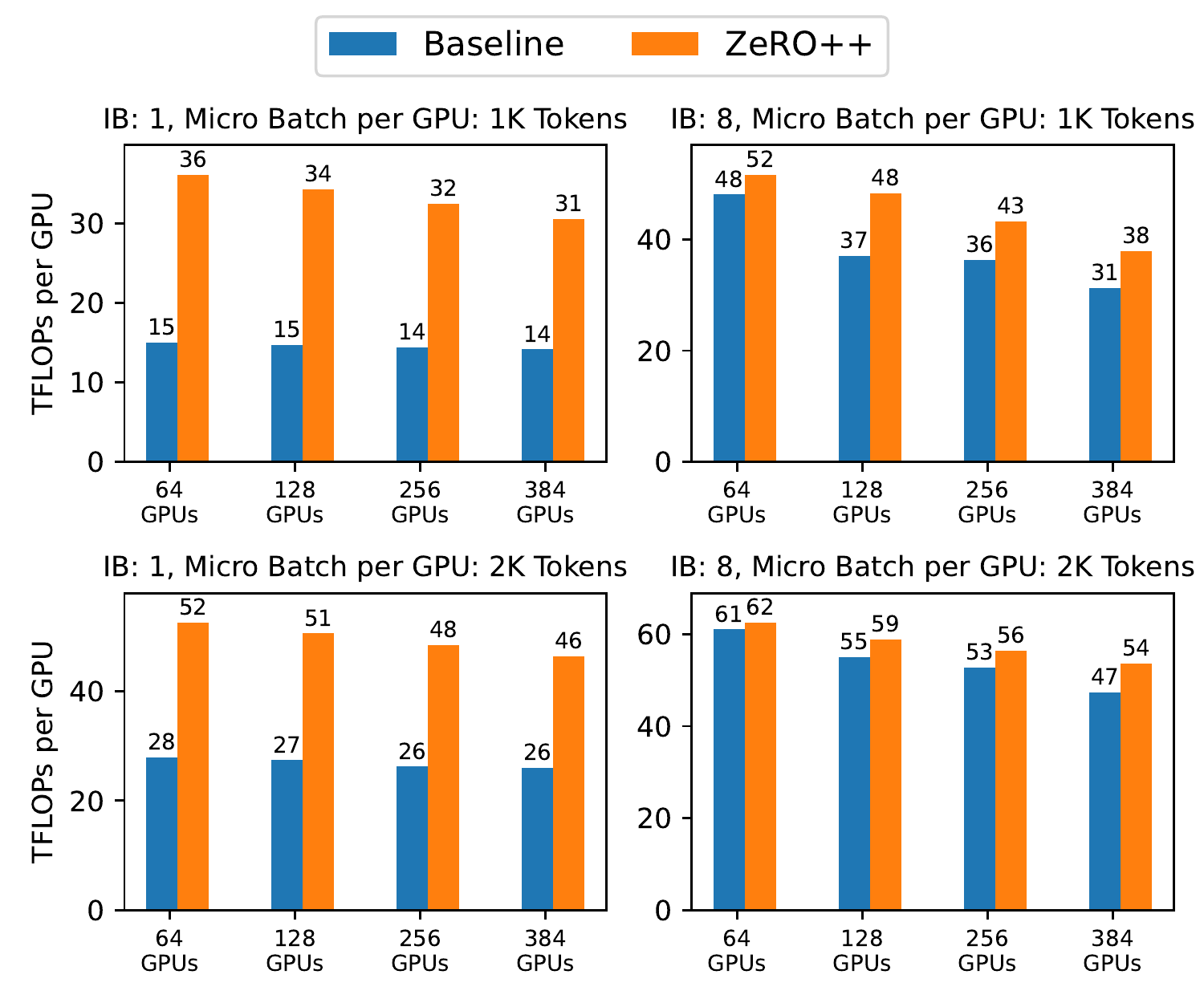}
    \caption{Scalability on up to 384 GPUs of 18B model with different numbers of InfiniBand connections and tokens per GPU}    \label{fig:eval_grid_plot}
\end{figure}

\subsection{E2E System Evaluations}
We evaluate ZeRO++ end-to-end performance here. One key metric we use here is the percentage of \emph{peak performance}, which is shown as equation~\ref{eq:peak-perf}.
\begin{equation}
peak\_performance = achieved\_TFLOPs / max\_TFLOPs
\label{eq:peak-perf}
\end{equation}

Given that we use V100 GPU, its $max\_TFLOPs$ is 120 TFLOPs~\cite{v100-tflop} for mixed precision computation. Thus, our reported \emph{peak performance} refers to the percentage number of $achieved\_TFLOPs / 120$.
\begin{table}[]
\centering
\caption{End-to-end speedup of ZeRO++ on 384 GPUs with different model sizes}
\label{tab:eval-384gpus}
\resizebox{\columnwidth}{!}{%
\begin{tabular}{cc|ccc|ccc}
\hline
     &    & \multicolumn{3}{c|}{1 IB Connection} & \multicolumn{3}{c}{8 IB Connections} \\ \hline
\begin{tabular}[c]{@{}c@{}}Model\\ Size\end{tabular} &
  \begin{tabular}[c]{@{}c@{}}Tokens\\ per GPU\end{tabular} &
  \begin{tabular}[c]{@{}c@{}}Baseline\\ TFLOPs\end{tabular} &
  \begin{tabular}[c]{@{}c@{}}ZeRO++\\ TFLOPs\end{tabular} &
  Speedup &
  \begin{tabular}[c]{@{}c@{}}Baseline\\ TFLOPs\end{tabular} &
  \begin{tabular}[c]{@{}c@{}}ZeRO++\\ TFLOPs\end{tabular} &
  Speedup \\ \hline
138B & 2K & 19.96      & 37.90      & 1.90x      & 47.55      & 55.30      & 1.16x      \\
138B & 1K & 11.25      & 21.81      & 1.94x      & 34.19      & 44.38      & 1.30x      \\ \hline
91B  & 2K & 19.99      & 38.06      & 1.90x      & 47.74      & 56.26      & 1.18x      \\
91B  & 1K & 11.27      & 21.93      & 1.95x      & 34.49      & 44.36      & 1.29x      \\ \hline
49B  & 2K & 20.06      & 38.08      & 1.90x      & 48.05      & 56.24      & 1.17x      \\
49B  & 1K & 11.27      & 21.95      & 1.95x      & 34.54      & 44.46      & 1.29x      \\ \hline
18B  & 2K & 25.98      & 46.40      & 1.79x      & 47.31      & 53.65      & 1.13x      \\
18B  & 1K & 14.15      & 30.57      & 2.16x      & 31.27      & 37.87      & 1.21x     
\end{tabular}%
}
\end{table}
\begin{figure}
    \centering
    \includegraphics[width=\columnwidth]{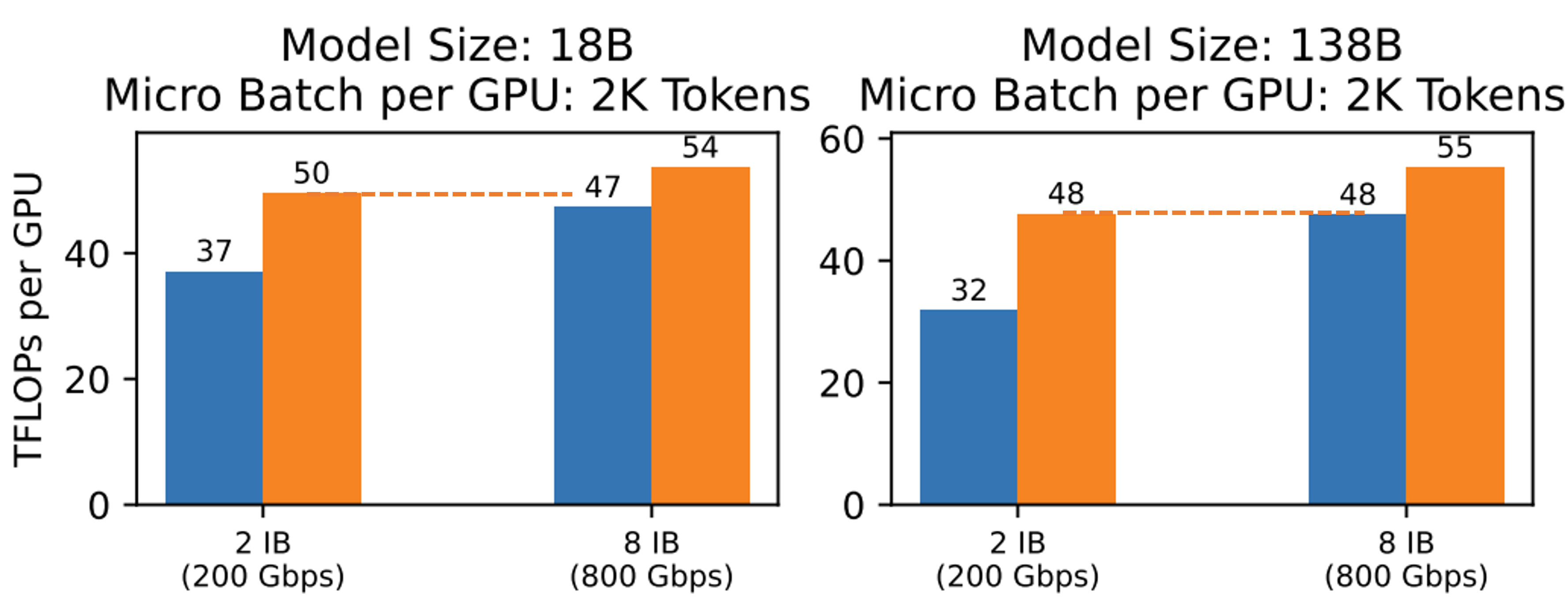}
    \caption{ZeRO++ achieving high bandwidth cluster performance with significantly lower bandwidth}    \label{fig:eval_ib_bandwidth_plot}
\end{figure}
\subsubsection{Scalability upto 384 GPUs}
In Figure \ref{fig:eval_grid_plot}, we present ZeRO++ scalability evaluation from 64 to 384 GPUs with 18B model on both low (1 IB) and high (8 IB) bandwidth clusters. On low bandwidth cluster, ZeRO++ achieves 30\% and 38.3\% of peak performance (120 TFLOPs) even at 384 GPUs for 1K and 2K batch sizes, which is much higher compared to 12.5\% and 21.6\% as baseline peak performance. This presents up to \textbf{2.4x} better throughput. On high bandwidth cluster, despite having significantly more bandwidth, ZeRO++ still enables up to 1.29x better throughput, and can achieve up 45\% of sustained peak throughput at 384 GPUs. ZeRO++ significantly speed up large scale training for low bandwidth clusters while archiving decent speedup even on high bandwidth clusters.
\subsubsection{Throughput for different model sizes}
Table \ref{tab:eval-384gpus} compares training throughput for models of 18B-138B on 384 GPUs between ZeRO++ and baseline on both low and high bandwidth clusters.
On low bandwidth cluster, ZeRO++ consistently achieves over 31.5\% and 18.1\% peak performance for 2K and 1K batch sizes on all models. Compared with the baseline peak performance of 16.6\% and 9.3\%, the speedup is up to \textbf{2.16x}. On high bandwidth cluster, ZeRO++ peak performances are 44.7\% and 31.5\%, which is 1.3x over the baseline peak performance of 31.5\% and 26.0\%. ZeRO++ is robust and offers consistent speedup across different model and batch sizes as well as across clusters with different network bandwidths. 

\subsubsection{Democratization for large scale training}
Figure \ref{fig:eval_ib_bandwidth_plot} compares the throughput of ZeRO++ on a low cross-node bandwidth (200 Gbps as 2 IB) cluster with the baseline running on 800 Gbps high-bandwidth (8 IB) cluster. 
For small model of 18B, ZeRO++ achieves a higher peak performance of 41.6\% compared with baseline peak performance of 39.1\% despite running with 4x lower cross-node bandwidth. For large model of 138B, ZeRO++ and baseline achieve the same peak performance of 40\%, but baseline runs at 4x higher cross-node bandwidth. 
This evaluation shows that ZeRO++ makes large scale training more accessible by significantly decreasing the minimum cross-node bandwidth requirement for efficient training. Furthermore, it demonstrates that optimized ZeRO++ implementation effectively translates the 4x communication reduction of ZeRO++ into real end-to-end system throughput gain.
%
%

\subsection{Throughput Breakdown and Analysis} 

\subsubsection{Impact of Individual Techniques}

\begin{figure}
    \centering
    \includegraphics[width=\columnwidth]{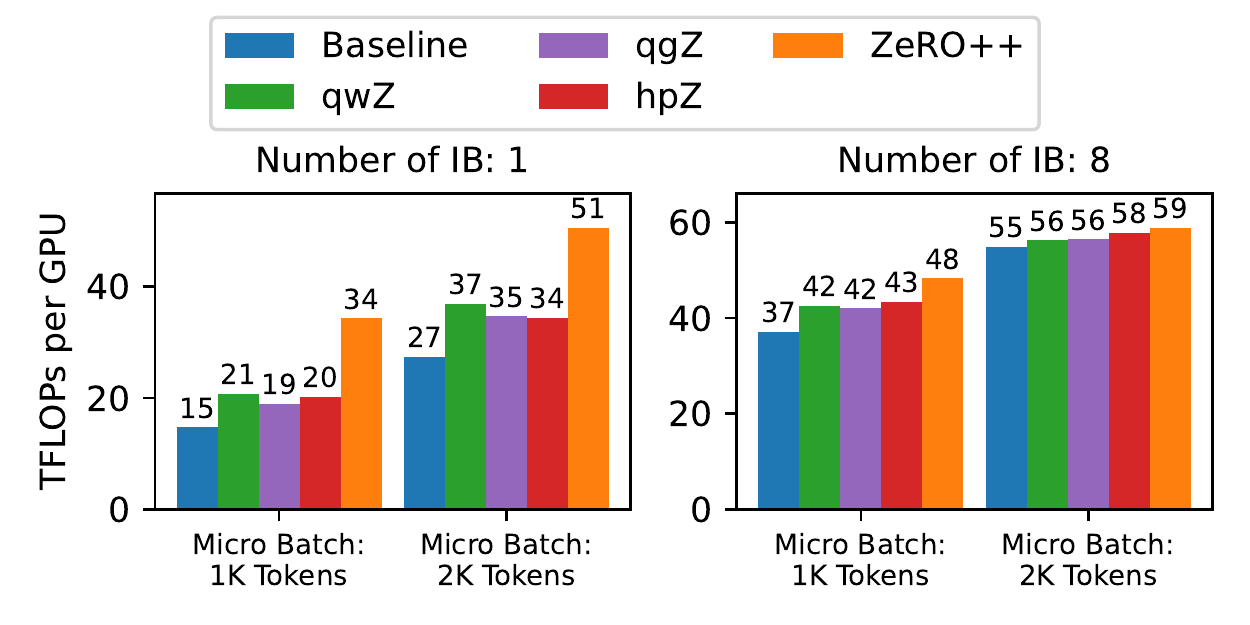}
    \caption{Throughput of 18B models on128 GPUs with ZeRO++, qwZ, qgZ, hpZ, and baseline on different numbers of InfiniBand connections}    \label{fig:eval_ablation_plot}
\end{figure}

In Figure \ref{fig:eval_ablation_plot}, we show the individual and combined impact of  qwZ, hpZ,  and qgZ, on the throughput of 18B model on 128 GPUs. On low bandwidth clusters, each of these techniques enables a speedup ranging from 1.3-1.4x compared with baseline, while achieving an aggregated speedup of up to 2.26x. 
Note that our TFLOPs throughput is calculated from wall-clock time measurement, ZeRO++ aggregated throughput gain is not equivalent to sum of qgZ, qwZ, hpZ gain.
We can validate the theoretical speedup with composition of our techniques by accumulating the speedup multiplicatively: $1.4*1.26*1.3=2.29$, which is very near to what we achieved as 2.26x.
 
For high bandwidth clusters, the individual speedup ranges between 1.13-1.16x, for a combined speedup of up to 1.3x. The figure demonstrates that each of these techniques has a similar impact towards throughput improvement and they compose effectively to produce a much larger aggregated speedup.

\subsubsection{Impact of Kernel Optimizations}


\begin{table}
\small
\centering
\caption{End-to-end performance when using ZeRO++ w.\textbackslash wo. optimized kernels}
\label{tab:eval-kernel}
\resizebox{0.8\columnwidth}{!}{%
\begin{tabular}{|c|c|c|c|}
\hline
 & \begin{tabular}[c]{@{}c@{}}Optimized \\ Quantization \\ Kernel\end{tabular} & \begin{tabular}[c]{@{}c@{}}Optimized\\ Fusion \\ Kernel\end{tabular} & TFLOPs \\ \hline
Baseline & N/A    & N/A    & 15   \\ \hline
ZeRO++   & No  & No  & 19.73 \\ \hline
ZeRO++   & No  & Yes & 21.6   \\ \hline
ZeRO++   & Yes & No  & 31.40 \\ \hline
ZeRO++   & Yes & Yes & \textbf{36.16}   \\ \hline
\end{tabular}%
}
\end{table}


Here, we evaluate the impact of our optimized kernels on ZeRO++ throughput using a 18B model running on 64 GPUs.


\noindent\textbf{Quantization Kernel: }
As shown in Table \ref{tab:eval-kernel}, compared with the baseline that uses PyTorch quantization~\cite{pytorch-quant}, our optimize quantization kernels can achieve up to 1.67x speedup in terms of end-to-end throughput. Also, the baseline implementation suffers performance degradation as group number increases which means the throughput gap will be larger when used with larger models.



\noindent\textbf{Kernel Fusion: }
As described in Section 4.2, kernel fusion is one of our key optimizations to improve memory throughput when executing sequences of CUDA kernels. Our fusion includes 1) tensor-reorder and quantization fusion 2) intra-node dequant, intra-node reduction and inter-node quant fusion. As shown in Table \ref{tab:eval-kernel}, we achieve up to  1.15x speedup on the end-to-end throughput.

\subsubsection{Comparing hpZ with MICS}
As previously discussed in Section ~\ref{sec:background}, closely related to hierarchical weight partition for ZeRO ($hpZ$) is $MiCS$\cite{zhang2022mics}. Key difference of the two methods is what data are replicated in secondary group; model weights are replicated in $hpZ$, entire model states are replicated in $MiCS$. 
Table \ref{tab:hpz-mic} shows per-GPU throughput of both methods for different model and token size configurations. The table also shows that given a secondary partition size of a single node (16 V100 GPUs), $hpZ$ can support 18 billion parameter model where as $MiCS$ reports out-of-memory (OOM) at this scale.
 
\begin{table}
\small
\centering
\caption{hpZ vs MiCS evaluation on a 4 node cluster (16 V100 GPUs per node)}
\label{tab:hpz-mic}
\resizebox{0.9\columnwidth}{!}{%
\begin{tabular}{|c|c|c|c|c|}
\hline Model Size
 & \begin{tabular}[c]{@{}c@{}} Token Size \end{tabular} & 
 \begin{tabular}[c]{@{}c@{}}ZeRO\\ TFLOPs \end{tabular} &
 \begin{tabular}[c]{@{}c@{}}hpZ\\ TFLOPs\end{tabular} &
 \begin{tabular}[c]{@{}c@{}}MiCS\\ TFLOPs \end{tabular} \\ \hline
7.5B & 1K     &  36.99 & 38.39   & 38.96   \\ \hline
7.5B   & 2K  & 53.3 & 54.4  & 52.72 \\ \hline
18B   & 1K  & 51.47 & 52.42 & OOM   \\ \hline
18B   & 2K & 60.94 & 61.44 & OOM \\ \hline
\end{tabular}%
}
\end{table}
\begin{figure}
    \centering
    \includegraphics[width=0.65\columnwidth]{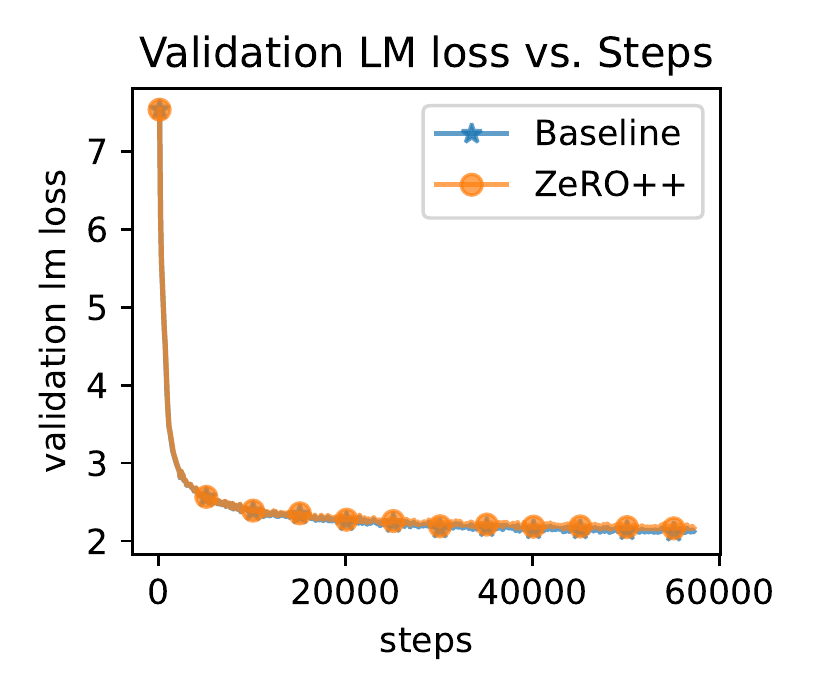}
    \caption{Training convergence for GPT-350M on 30B tokens}    \label{fig:eval_convergence}
\end{figure}

\subsection{Model convergence analysis}

Next we evaluate ZeRO++'s impact on model convergence by training GPT-350M model with 30B tokens on the pile dataset~\cite{biderman2022datasheet} using ZeRO++, ZeRO++ with basic (non-blocked) quantization, and ZeRO-3 as baseline. All hyperparameters are kept the same between baseline training and ZeRO++ trainings to ensure fair comparison. The convergence is measured by the validation LM loss.

As shown in Figure \ref{fig:eval_convergence}, we present end-to-end training trace.
The training with basic (non-blocked) quantization diverged at the beginning so there is no visible data, on the contrary,  ZeRO++ is closely aligned with the baseline, and also confirms our previous analysis of better quantization accuracy by using block based quantization.

\begin{table}[]
\centering
\caption{Validation loss at the end of training (GPT 350M / 30B tokens)}
\label{tab:eval_convergence}
\resizebox{0.8\columnwidth}{!}{%
\begin{tabular}{|c|c|}
\hline
 & Evaluation LM loss \\ \hline
Baseline & 2.121762 \\ \hline
\begin{tabular}[c]{@{}c@{}}ZeRO++\\ (hpZ\&qwZ\&qgZ on)\end{tabular} & 2.165584 \\ \hline
\begin{tabular}[c]{@{}c@{}}ZeRO++\\ (hpZ\&qwZ on;\\ qgZ on for first 50\%)\end{tabular} & 2.134013 \\ \hline
\begin{tabular}[c]{@{}c@{}}ZeRO++\\ (hpZ\&qwZ on; qgZ off)\end{tabular} & 2.121653 \\ \hline
\end{tabular}%
}
\end{table}

We further extended the convergence evaluation by comparing the final evaluation loss at the end of training. As shown in Table \ref{tab:eval_convergence}, even with all three optimizations on, the final evaluation loss is only off by 1\%. We further merged this convergence gap by using a straightforward interleaving schedule where the hierarchical partitioning and quantized weights are turned on throughout the training and the quantized gradient is only turned on for the first 50\% of training. For a more extended case, we also evaluate hierarchical partitioning and quantized weights alone. The results suggest our convergence is identical to the baseline in this case.







\section{Conclusion}

This paper present ZeRO++, an efficient collective communication solution for giant model training using ZeRO stage-3. We optimize both model weights and gradients communication in forward and backward pass of each training iteration. To reduce communication volume of model weights in forward propagation, we adopt block-based quantization and data pre-fetching. To remove cross-node communication of weights during backward pass, we hold secondary model partition on each node to trade memory for communication. To minimize gradient communication during backward propagation, we design and implement a novel all-to-all based gradient quantization and reduction scheme. By incorporating all the three optimizations above, we improve system throughput up to 2.16x in large scale model training using 384 V100 GPUs. 
We envision ZeRO++ as the next generation of easy-to-use framework for training giant models at trillion-level model scale. 






\section{Authorship and Major Credit Attribution}

\begin{itemize}
\item \textbf{Guanhua Wang:} design and implementation of qgZ, code integration, high performance quantization kernels design and implementation, solving all CUDA kernel conflicts in code merging, majority of paper writing. 
\item \textbf{Heyang Qin:} design and implementation of qwZ, code integration/resolve conflicts in code merging, experimental design and evaluation, in depth convergence study.
\item \textbf{Sam Ade Jacobs:} design and implementation of hpZ, code integration/resolve conflicts in code merging.  
\item \textbf{Connor Holmes:} design and implementation of high performance quantization kernels.
\item \textbf{Samyam Rajbhandari:} chief architect 
\item \textbf{Olatunji Ruwase:} technical support
\item \textbf{Yuxiong He:} team lead

\end{itemize}

\clearpage
\bibliographystyle{ACM-Reference-Format}
\bibliography{sample}

\end{document}